\documentclass{ifacconf}

\usepackage{graphicx}      % include this line if your document contains figures
\usepackage{natbib}        % required for bibliography
\usepackage{comment}
\usepackage{amsmath}
\usepackage{caption}
\usepackage{subcaption} 
\usepackage{url}
\usepackage{xcolor}
\usepackage{array}
\DeclareMathOperator*{\argmin}{arg\,min}

\begin{document}
\begin{frontmatter}
\title{Joint Travel Route Optimization Framework for Platooning\thanksref{footnoteinfo}\thanksref{copyrightnotice}} 

\thanks[footnoteinfo]{This work is financed by the European Union—NextGenerationEU (National Sustainable Mobility Center CN00000023, Italian Ministry of University and Research Decree n. 1033—17/06/2022, Spoke 9).}
\thanks[copyrightnotice]{© 2025 the authors. This work has been accepted to IFAC for publication under a Creative Commons Licence CC-BY-NC-ND.}

\author{Akif Adas,} 
\author{Stefano Arrigoni,}
\author{Mattia Brambilla,}
\author{Monica Barbara Nicoli,} 
\author{Edoardo Sabbioni}
 
\address{Politecnico di Milano, 
   Milan, 20156 Italy \\ (e-mail: akif.adas@polimi.it).}

\begin{abstract}

Platooning represents an advanced driving technology designed to assist drivers in traffic convoys of varying lengths, enhancing road safety, reducing driver fatigue, and improving fuel efficiency. 
Sophisticated automated driving assistance systems have facilitated this innovation. 
Recent advancements in platooning emphasize cooperative mechanisms within both centralized and decentralized architectures enabled by vehicular communication technologies.
This study introduces a cooperative route planning optimization framework aimed at promoting the adoption of platooning through a centralized platoon formation strategy at the system level.
This approach is envisioned as a transitional phase from individual (ego) driving to fully collaborative driving. 
Additionally, this research formulates and incorporates travel cost metrics related to fuel consumption, driver fatigue, and travel time, considering regulatory constraints on consecutive driving durations.
The performance of these cost metrics has been evaluated using Dijkstra’s and A* shortest path algorithms within a network graph framework. 
The results indicate that the proposed architecture achieves an average cost improvement of 14\% compared to individual route planning for long road trips.

\end{abstract}

\end{frontmatter}
%=====================================================

\section{Introduction}

%\textcolor{red}{Add and comment this ref \cite{10132885}}

Traffic accidents on highways are of essential concern due to their fatal results, with various contributing factors identified in numerous studies. 
Fatigue driving and dozing at the wheel are major causes, as stated by \cite{ha2}. % , ha1
Some of these causes are avoided by the advancements in automated driving systems such as adaptive cruise control, emergency braking, lane keeping assistance and driver monitoring systems, e.g., eye tracking, drowsiness detection, and warning using facial and hand gestures (\cite{driver_monitoring}). 
% ACC (\cite{ACC})
% EB (\cite{EB})
% LKA (\cite{lka})
% ET (\cite{eye_tracking})
% DD (\cite{drawsiness})
% HG (\cite{driver_monitoring})
These systems increase the safety of passengers traveling on highways and road capacity by allowing vehicles to cruise closer, thanks to reduced reaction times. 

In automated driving systems, vehicles and roadside infrastructure employ imaging sensors, e.g., camera, lidar, and radar, to sense the environment and extract crucial road information such as the distance between vehicles or any hazards from side, e.g., pedestrians 
(\cite{OJITS25}). 
Although these systems are capable of assisting human drivers in their tasks, they cannot currently guarantee an adequate level of safety to properly achieve level 4 or higher as prescribed by \cite{SAE}. As a promising and more easily achievable solution, platooning has gained immense attention in academia and original equipment manufacturers (OEMs).  

\cite{platooning_def} defines platooning as an emerging driving strategy where multiple vehicles travel collaboratively as a single string. 
Platooning is expected to provide improved fuel efficiency, higher traffic capacity, reduced traffic congestion, and fewer traffic accidents thanks to reduced air drag, closer driving distances, and leveraged awareness through vehicle-to-everything (V2X) communications (\cite{V2X_akif} and \cite{V2X_poli}).
These improvements have been well studied and reported by \cite{platooning_eff1} and \cite{platooning_eff2}.
In order to assign a vehicle to a platoon, vehicle formation is dealt with by either centralized or decentralized algorithms, as emphasized in \cite{platoon_formation}. While centralized model predictive control (MPC) solutions focus on centralizing decision-making to optimize the interaction between vehicles, decentralized MPC and deep reinforcement-based solutions propose scalable platoon string formation at the cost of lower efficiency.
 
Recent research has focused on the challenges related to platooning. 
\cite{plat_csec_2} and \cite{plat_csec_3} proposed solutions for platooning cybersecurity. %\cite{plat_csec_1},
Platoon management approaches have been studied by 
%\cite{dsrc_platoon}, \cite{plat_man_1}, 
%\cite{plat_csec_2}, 
\cite{plat_man_3} and \cite{plat_man_4}. Besides,  \cite{plat_comm_1} and \cite{plat_comm_2} highlighted the effective communication methods of platooning. 

Moreover, various projects have explored platooning systems in the real world. 
Some examples are SARTRE (Europe), PATH (USA), and Energy ITS (Japan) as reported by \cite{platooning_practices}.
The EU ENSEMBLE Project focuses on developing and implementing multi-brand truck platooning solutions in multi-brand settings \cite{10132885}.
%To the best of the authors' knowledge, these studies strongly assume that vehicles are ready for collaboration on highways. 
%Nevertheless, at the time of writing this paper, a few vehicles have sophisticated cooperation and driving technologies. As a consequence, engaging in platoon formation is limited to controlled research experiments in restricted environments. 

\textbf{Contribution.} This paper 
%proposes a novel
introduces a preliminary version of a
framework for optimizing long-distance route planning via a centralized system that leverages V2X communications, enabling vehicles to form platoons at the earliest and most feasible opportunity. 
%To this purpose, it utilizes widely adopted algorithms such as brute-force search and Dijkstra's algorithm, considering their seamless integration into existing systems.
In the proposed approach, vehicles can coordinate with other road participants to schedule long-distance routes with shared destinations or similar paths, which optimizes not only their individual travel costs but also enhances overall traffic efficiency. This joint optimization ensures that vehicles find the most advantageous platoon compositions based on parameters like journey time, fuel economy, and driver fatigue. Through this collaborative model, the system seeks to redefine route planning by prioritizing network-wide efficiency and cost-effectiveness, ultimately paving the way for scalable, cooperative transportation ecosystems. 

Besides, this study defines different routing cost functions and examines their impacts on route planning optimization by leveraging Dijkstra's and A* algorithms. 

The remainder of the paper is structured as follows: Sec.~\ref{Sec:TRP} presents the route optimization solutions and elaborates on different cost term definitions; Sec.~\ref{Sec:JRO} introduces the developed solution; Sec.~\ref{Sec:Res} outlines numerical experiments and achieved improvements in terms of driver fatigue level, travel time and fuel consumption. Finally, Sec.~\ref{Sec:Conc} concludes the paper and portrays future work. 
\section{Travel Route Planning}
\label{Sec:TRP}

\subsection{Problem statement}
\label{Sec:PS}

\begin{comment}
\begin{itemize}
    \item explain A* algorithm and define heuristic carefully
\end{itemize}
\end{comment}

A directed, finite graph $G=(V, E)$ represents a route, where $V$ is the set of vertices (intersections) and $E$ is the set of directed edges (road segments). A vehicle route is defined by a sequence of vertices, determined by selecting optimal edges. Examples of algorithms for route identification include Dijkstra and A*, which are described in the following.

\textbf{\textit{Dijkstra's algorithm}} (\cite{dijkstra}). 
It is a foundational graph traversal algorithm designed to determine the shortest path from a source vertex $s$ in a weighted graph $G$ with non-negative and additive edge weights $w(u,v)$. 
It maintains a set $S$ of vertices with known shortest paths and a distance array $d$, where $d[s]=0$ and $d[v]=\infty$ for $v \in V \setminus \{s\}$.

The algorithm iteratively selects $u \in V \setminus S$ with minimal $d[u]$, adds $u$ to $S$, and updates distances of neighbors $v$:

\begin{equation}
    d[v] = \min(d[v], d[u] + w(u,v)).
\end{equation}

Termination occurs when $S=V$, yielding $d[v]$ as the shortest path distance. The algorithm, a greedy approach, achieves $O(|E|+|V|\log|V|)$ complexity with a priority queue.

\textbf{\textit{A* algorithm}} (\cite{Astar}). It is an extension of Dijkstra's, aiming to find the shortest path from a source vertex $s$ to a goal vertex $g$ in a weighted graph $G$ with non-negative edge weights $w(u,v)$. It introduces a heuristic function $h(v)$, estimating the cost from vertex $v$ to $g$.
A* maintains two sets: an open set $O$ of vertices to be explored, and a closed set $C$ of explored vertices. Each vertex $v$ is associated with a cost $f(v) = g(v) + h(v)$, where $g(v)$ is the accumulated cost from $s$ to $v$. Initially, $O=\{s\}$ and $C=\emptyset$.

The algorithm iteratively selects $u \in O$ with minimal $f(u)$, adds $u$ to $C$, and removes $u$ from $O$. If $u=g$, the shortest path is found. For each neighbor $v$ of $u$, the algorithm computes a tentative $g(v)$ via $u$ as follows:

\begin{equation}
    g'(v) = g(u) + w(u,v).
\end{equation}

If $v \notin O \cup C$ or $g'(v) < g(v)$, $g(v)$ is updated, $v$'s parent is set to $u$, and $v$ is added to $O$ if not already present.

Given a weighted graph $G=(V,E)$ with edge weights $w(u,v)$, the cost of a path $\Gamma$ from a source vertex $i$ to a destination vertex $j$, denoted by $\mathcal{C}_\Gamma$, is defined as the cumulative sum of the weights of the edges traversed in $\Gamma$. Formally,

\begin{equation}
  \mathcal{C}_\Gamma = \sum_{(u,v) \in \Gamma} w(u,v),  
\end{equation}

where $\Gamma$ is a sequence of vertices and edges representing a path such that $\Gamma \in \mathcal{P}_{i \rightarrow j}$. Here, $\mathcal{P}_{i \rightarrow j}$ represents the set of all possible paths from vertex $i$ to vertex $j$.

%Termination occurs when $O$ is empty or $g$ is reached. A* is optimal if $h(v)$ is admissible (never overestimates the cost) and consistent (obeys the triangle inequality). The complexity depends on the heuristic's quality. the Big O notation of the A* algorithm is not a single, fixed value. It's highly dependent on the heuristic function used. 

\begin{comment}
To ensure the optimality of A*, three criteria must be satisfied: locally finite, admissibility, and monotonicity.    
A locally finite graph is defined as a graph in which each node has a finite branching factor. This means that no node has an infinite number of neighbouring nodes to expand, thus ensuring paths do not branch indefinitely. 
A node's branching factor represents the number of immediate successors or child nodes that can be generated from it.
An admissible heuristic is one that is consistently optimistic; it either accurately estimates or underestimates the cost from the current node to the goal. 
Importantly, an admissible heuristic never overestimates the actual cost of reaching the goal.
\end{comment}

\subsection{Travel Cost Definitions}
\label{Sec:Travel Cost}

The edge weight $w(u,v)$ for $(u,v) \in E$ is defined as a combination of cost terms. Specifically, we consider travel time, distance, fuel consumption, and fatigue level.

%The cost $g_{e}$ of each edge can be defined by accounting for different contributions: in this paper, three different cost terms (travel time, fuel consumption and fatigue level) are proposed and individually used. Their description is as follows.

\textit{\textbf{Travel Time}}. 
Consider a set of vehicles $V$, where a subset $V_p \subseteq V$ forms a platoon on a highway after initiating driving within a city. 
Assume all vehicles $v \in V$ maintain a constant speed $v_{c} = 110$ km/h.

We posit that standard driving time regulations, exemplified by the European limit of $T_{EU} = 32400$ s (9 hours) with a mandatory rest period of $T_{r} = 2700$ s (45 minutes) (\cite{consecutiveDriving}), are not applicable to platoon members $v \in V_p$. This exemption is predicated on the assumption that drivers within a platoon are relieved of active driving responsibilities.

To facilitate a comparative analysis that avoids the inherent bias of comparing a platoon vehicle with a fatigued non-platoon vehicle, we distribute the rest time $T_{r}$ proportionally to the traversed distance along the path $\Gamma$. This methodology allows platoon members to exceed the $T_{EU}$ threshold through distributed rest periods.

The travel cost $\mathcal{C}_{T}^{(I)}$ for a non-platoon vehicle $v \in V \setminus V_p$ is defined as:

\begin{equation}
\mathcal{C}_{T}^{(I)} = \sum_{(u,v) \in \Gamma} \frac{d(u,v)}{v_{c}} \left(1 + \frac{T_{r}}{T_{EU}}\right),
\end{equation}

where $d(u,v)$ (in meters) represents the distance of the edge $(u,v) \in \Gamma$.

The travel cost $\mathcal{C}_{T}^{(P)}$ for a platoon member $v \in V_p$ is computed as:

\begin{equation}
\mathcal{C}_{T}^{(P)} = \sum_{(u,v) \in \Gamma} \frac{d(u,v)}{v_{c}}.
\end{equation}

\textit{\textbf{Fuel consumption.}} 
Fuel consumption, denoted as $F$, is determined as a function of the traversed distance. According to \cite{plat_fuel_consumption}, vehicles $v \in V_p$ operating within a platoon experience aerodynamic benefits, resulting in fuel economy improvements ranging from 3\% to 18\%. This improvement is contingent upon the vehicle's position within the platoon and the prevailing operational conditions. Specifically, the lead vehicle exhibits minimal fuel savings, while trailing vehicles achieve substantial reductions due to diminished air resistance. This phenomenon underscores the potential of cooperative driving technologies to enhance energy efficiency and mitigate emissions within vehicular networks. Cost due to fuel consumption is denoted $C^{(I)}_{FC}$ and $C^{(P)}_{FC}$ for individual and platoon driving, respectively.

%It is calculated based on the traveled distance. According to \cite{plat_fuel_consumption}, vehicles operating within a platoon benefit from aerodynamic advantages, leading to fuel economy improvements ranging from 3\% to 18\%, depending on the vehicle's position in the platoon and the specific operation conditions. The lead vehicle experiences minimal benefits while trailing vehicles achieve significant reductions in fuel consumption due to reduced air resistance. These savings highlight the potential of cooperative driving technologies in enhancing energy efficiency and reducing overall emissions in vehicular networks. 
%In this work, the fuel consumption improvement was considered 18\% for the successor vehicles.

\textit{\textbf{Fatigue level.}} 
Driver fatigue represents a significant consequence of prolonged vehicular operation, with substantial implications for traffic safety and individual health. \cite{fatigueRelatedAccidents} report that fatigue, both directly and indirectly, contributes to 30\%-40\% of traffic accidents.

To quantify driver fatigue, researchers have employed both medical instrumentation and subjective assessments. For instance, \cite{driverFatigue} conducted a series of experiments involving 19 subjects of mixed genders. These experiments yielded Karolinska Sleepiness Scale (KSS) scores, which were subsequently transformed into continuous fatigue values using cubic spline interpolation.

In their model, Zhang et al. (2019) defined fatigue as a cumulative function of three primary factors: the temporal influence of circadian rhythms, the duration of consecutive driving, and the quality of prior sleep. Mathematically, this cumulative fatigue was modeled as:

%It q%is a prominent consequence of driving, along with other health-related implications. Fatigue, directly and indirectly, impacts 30\%-40\% traffic accidents as reported by \cite{fatigueRelatedAccidents}. Some research has been done by assessing the fatigue level with medical instruments. 
%\cite{driverFatigue} conducted several campaigns with 19 subjects of mixed genders. As a result of all experiments, they reported the Karolinska Sleepiness Scale (KSS) and converted it into consecutive fatigue values by cubic spline interpolation. 
%See Table II in the paper for the relation between KSS and fatigue values. 
%They defined fatigue as the accumulation of three factors: the time range of circadian rhythms, consecutive driving time, and sleeping time. 
%This paper solely focuses on route optimization. Hence, only the impact of consecutive driving time has been taken into account, as the other components depend on the drivers' habits. Fatigue level is formulated for morning, afternoon and night driving separately. For the sake of simplicity, this paper assumes that joint route optimization is employed only for morning driving. 
%\cite{driverFatigue} modeled the fatigue as: %$\text{F}^{(\text{M})}$, and afternoon, $\text{F}^{(\text{A})}$, are unitless and

\begin{align}
F_{t_{dm}} &= \alpha_1 e^{-\left(\frac{t_{dm} - \beta_1}{\epsilon_1}\right)^2} \label{eq:Ftdm} \\
F_{t_{da}} &= \alpha_2 e^{-\left(\frac{t_{da} - \beta_2}{\epsilon_2}\right)^2} + \alpha_3 e^{-\left(\frac{t_{da} - \beta_3}{\epsilon_3}\right)^2} \label{eq:Ftda} \\
F_{t_{dn}} &= \alpha_4 e^{-\left(\frac{t_{dn} - \beta_4}{\epsilon_4}\right)^2} + \alpha_5 e^{-\left(\frac{t_{dn} - \beta_5}{\epsilon_5}\right)^2} \nonumber \\
        &+ \alpha_6 e^{-\left(\frac{t_{dn} - \beta_6}{\epsilon_6}\right)^2} \label{eq:Ftdn}
\end{align}

where $t_{dm}$, $t_{da}$, and $t_{dn}$ are the driving times in the morning, afternoon and night, respectively. The coefficients are given in Table \ref{tab:fatigue_coefficients}. The fatigue values are aggregated to find the overall fatigue determined as:

\begin{equation}
    C^{(I)}_F = F = F_{tdm} + F_{tda} + F_{tdn} ,
    \label{eq:cost_fatigue}
\end{equation}

\begin{table}[!t]
\centering
\caption{Coefficients for fatigue computation}
\label{tab:fatigue_coefficients}
\renewcommand{\arraystretch}{1} % Adjust row height for better readability
\begin{tabular}{ >  {\centering\arraybackslash}m{1.5cm} | >{\centering\arraybackslash}m{1.5cm} | >{\centering\arraybackslash}m{1.5cm} | >{\centering\arraybackslash}m{1.5cm}}
\hline
Symbol & Value & Symbol & Value \\
\hline
$\alpha_1$ & 60.83 & $\beta_1$ & 8834 \\
$\epsilon_1$ & 4760 & $\alpha_2$ & 22.1 \\
$\beta_2$ & 9675 & $\epsilon_2$ & 6142 \\
$\alpha_3$ & 92.1 & $\beta_3$ & $1.382 \times 10^4$ \\
$\epsilon_3$ & 6358 & $\alpha_4$ & 2.599 \\
$\beta_4$ & 5046 & $\epsilon_4$ & 1257 \\
$\alpha_5$ & 92.1 & $\beta_5$ & $1.382 \times 10^4$ \\
$\epsilon_5$ & 6358 & $\alpha_6$ & 22.1 \\
$\beta_6$ & 9675 & $\epsilon_6$ & 6142 \\
\hline
\end{tabular}
\end{table}

%The overall cost due to fatigue is evaluated as follows:
%\begin{equation}
%    \mathcal{C}_{\Gamma,f}^{I} = \text{F}^{(\text{M})}(t_\Gamma),
%\end{equation}
%\textcolor{red}{Unclear equation} 
%where $t_\Gamma$ is the time to complete the entire route.

where fatigue cost of individually driving is denoted as $C^{(I)}_F$. Platooning eases the driving process by reducing the responsibilities of drivers. 
Member vehicles of a platoon string are assumed to perform fully autonomous driving by communicating with the master vehicle and the other members and by enhanced environmental perception capabilities. 
%After a platoon is formed, the driver of the member vehicle is presumed to perform fatigue-free driving. 

Equations \eqref{eq:Ftdm}, \eqref{eq:Ftda}, and \eqref{eq:Ftdn} are nonlinear and non-additive. 
Hence, \eqref{eq:cost_fatigue} violates the optimality of Dijkstra's algorithm. 
Considering this, fatigue is incorporated into A* as a heuristic.
Here, the heuristic is artificially inflated to make A* more greedy and formulated as follows:

\begin{equation}
     h(v) = \varphi C^{(I)}_{\Gamma, M} C^{(I)}_F
\end{equation}

where $\varphi = 96.06$ and $C^{(I)}_{\Gamma, M}$ is the travel cost of master vehicle. The computation of $C^{(I)}_F$ relies upon an approximation of the travel time to the destination, thereby ensuring the property of monotonicity. Nevertheless, the potential for $h(v)$ to dominate $f(v)$ leads to transforming the A* search algorithm into a greedy search paradigm. 

\textit{\textbf{Overall Cost Term.}} 
Edge weights $w(u,v)$ are calculated as a mixture of the aforementioned cost terms. $C^{(P)}_{T}$, $C^{(I)}_{T}$, $C^{(I)}_{FC}$ and $C^{(P)}_{FC}$ are rescaled to $d(u,v)$ with coefficients $\kappa^{(P)}_{T}$, $\kappa^{(I)}_{T}$, $\kappa^{(I)}_{FC}$ and $\kappa^{(P)}_{FC}$, respectively. In turn, the cost function is formulated as:

\begin{equation}
  \mathcal{C}^{(I)}_{\Gamma, O} = d(u,v) + \kappa^{(I)}_{T} C^{(I)}_{T} + \kappa^{(I)}_{FC} C^{(I)}_{FC},  
\end{equation}

and the cost function with platooning is formulated as:

\begin{equation}
  \mathcal{C}^{(P)}_{\Gamma, O} = d(u,v) + \tau \,\kappa^{(P)}_{T} C^{(P)}_{T} + \xi \,\kappa^{(P)}_{FC} C^{(P)}_{FC},  
\end{equation}

where $\tau \in [0,1]$ and $\xi\in [0,1]$ are mixing rates. These variables are used to determine the impact of platooning on route planning. 
Eventually, the journey cost is obtained by aggregating cost during the individual route, $\Gamma^{(I)}$, and the platoon route, $\Gamma^{(P)}$ as:

\begin{equation}
  \mathcal{C}_\Gamma = \sum_{(u,v) \in \Gamma^{(I)}} \mathcal{C}^{(I)}_{\Gamma, O} + \sum_{(u,v) \in \Gamma^{(P)}} \mathcal{C}^{(P)}_{\Gamma, O} .
\end{equation}

%---------------------------------- 
\section{Joint Route Optimization}
\label{Sec:JRO}

\begin{figure} [!t]
    \centering
    \includegraphics[width=0.78\linewidth]{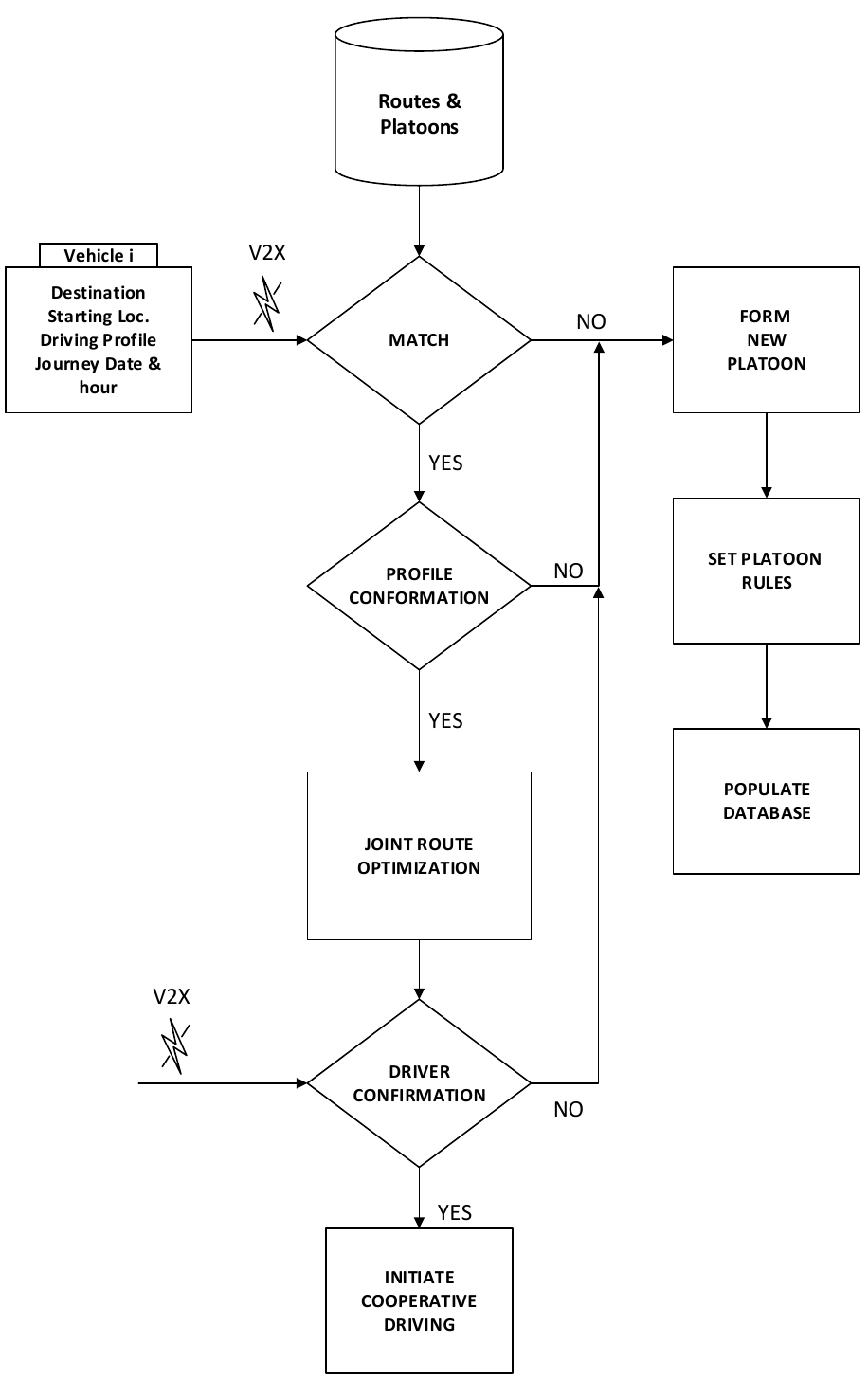}
    \caption{Joint route optimization system diagram.}
    \label{fig:block_diagram}
\end{figure}

\begin{figure*}[!tb]
    \centering
    \includegraphics[height=6cm]{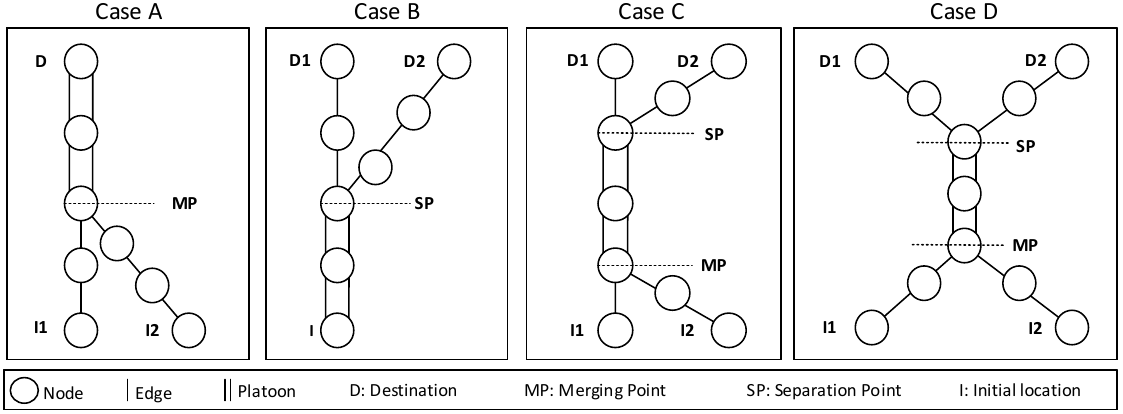}
    \caption{Cases of joint route optimization.}
    \label{fig:joint_root_optimization_cases}
\end{figure*}

This section delineates the proposed route optimization methodology and categorizes relevant driving scenarios. The objective of the optimization is to determine optimal platoon formations and maximize platoon duration, denoted as $\tau_p$, subject to the constraint of minimizing overall journey cost. This approach offers significant advantages over individual vehicle operations, particularly for extended journeys. We assume vehicles are equipped with Level-2 autonomous driving capabilities on highways and implement a periodic rotation of the platoon leader.

The vehicular network comprises $N_v$ members, facilitating communication with a centralized platoon planner via Vehicle-to-Everything (V2X) technology. As illustrated in Fig. 1, the optimizer receives input consisting of vehicle locations, destinations, and user-defined driving profiles. These profiles encompass parameters such as maximum and average driving speed, consecutive driving time limits, and driver preferences.

Initially, the route planner queries a database for common routes. If no matching route is found, the querying vehicle assumes the role of the master, generating a new route and establishing platooning protocols. Conversely, the vehicle possessing the longest individual route estimate, denoted as $b_\Gamma$, is designated as the master. This route serves as the reference for the remaining network members. Upon confirmation of driving profile compatibility, the joint route optimization algorithm is executed, and the driver is presented with the proposed route. It is noted that this route may exceed the length of the individually calculated route. Driver confirmation triggers the initiation of a cooperative driving procedure.

The multi-vehicle joint route optimization problem is addressed for a master vehicle and a member vehicle across four distinct scenarios, as depicted in Fig. 2. The route taken by a member vehicle is referred to as a member route.

\begin{itemize}
    \item \textbf{Case A:} A member route merges with the master route at a merging point (MP), and the vehicles proceed as a platoon for the remaining journey.
    \item \textbf{Case B:} Vehicles initiate their journey as a platoon and separate at a separation point (SP).
    \item \textbf{Case C:} A member vehicle travels within the platoon between a merging point (MP) and a separation point (SP).
    \item \textbf{Case D:} Both routes are optimized concurrently, transitioning from a master-member hierarchy to a member-member relationship. This scenario will be explored in future work.
\end{itemize}

\section{Numerical Results}
\label{Sec:Res}

\begin{table}[!t]
\centering
\caption{Simulation parameters}
\label{tab:simulationParams}
\renewcommand{\arraystretch}{1} % Adjust row height for better readability
\begin{tabular}{ >  {\centering\arraybackslash}m{4cm} | >{\centering\arraybackslash}m{4cm} 
}
\hline
Description & Value  \\
\hline
Area $[X, Y]$ boundaries & [1e6 1e6] m  \\
Number of nodes & 100  \\
Number of edges & 500  \\
Edge dropout rate & 0.2 \\
Spawn circle diameter & 1e3 m  \\
Minimum route length & 5e5 m  \\
Number of vehicles $(N_v)$ & 10  \\
Monte Carlo iterations & 100 \\

\hline
\end{tabular}
\end{table}

This section provides the performance analysis of the proposed platooning route optimization strategy, focusing on evaluating fuel efficiency, consecutive travel time, and drivers' fatigue. 
The conducted analyses refer to case C, i.e., a condition where both merging and separation operations apply. 

\begin{figure}
    \centering
    \includegraphics[width=1\linewidth]{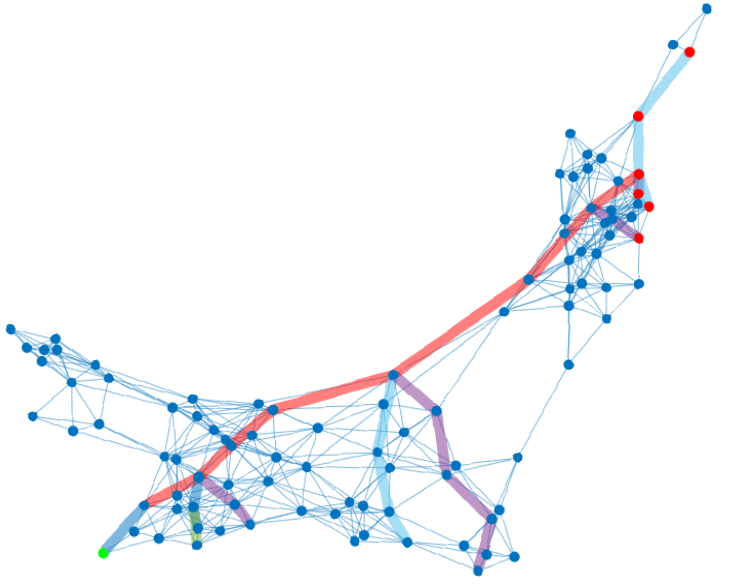}
    \caption{Example of road graph with the longest platooning path highlighted with red, spawn points with red, and member vehicle routes with various colors.}
    \label{fig:networkGraph}
\end{figure}

The network graph is randomly generated using the parameters given in Table \ref{tab:simulationParams}. The network spans \textit{Area [X, Y] boundaries} and is composed of \textit{Number of nodes} nodes. The fully connected network is firstly pruned to \textit{Number of edges} edges and secondly, edges randomly discarded according to \textit{Edge dropout rate}. Member vehicles are spawned in a circle centered at a randomly chosen master vehicle spawn point with a diameter of \textit{Spawn circle diameter}. Vehicles are obligated to travel \textit{Minimum route length}. The network is altered for \textit{Monte Carlo iterations}. An example graph network is visualized in Fig. \ref{fig:networkGraph} in which junctions are colored blue and the platoon red. In addition, member vehicle spawn locations are highlighted in red.      

\begin{figure}
    \centering    \includegraphics[width=1\linewidth]{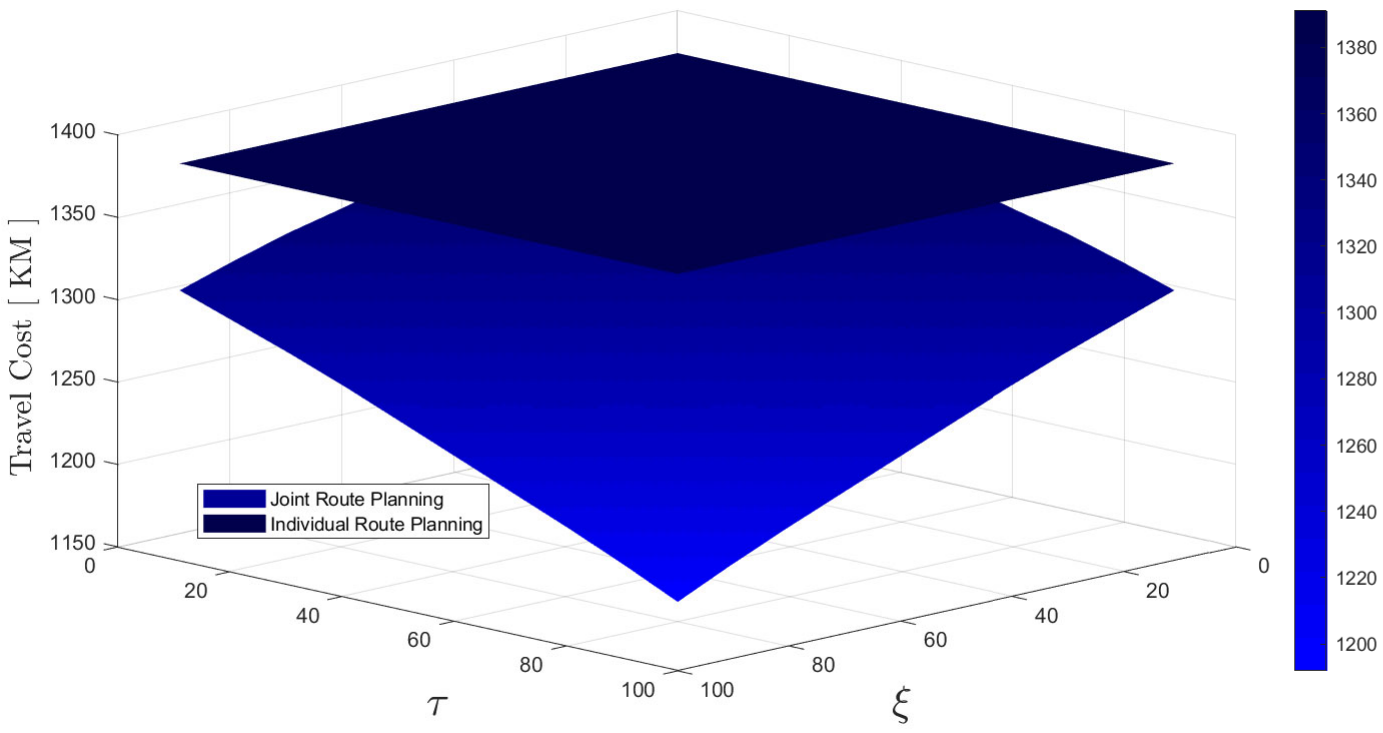}
    \caption{Average travel cost for individual and joint route planning using Dijkstra's algorithms for varying fuel consumption and travel time gains. Mixing terms represent percentages.}
    \label{fig:surfPlot}
\end{figure}

\begin{figure}
    \centering
    \includegraphics[width=1\linewidth]{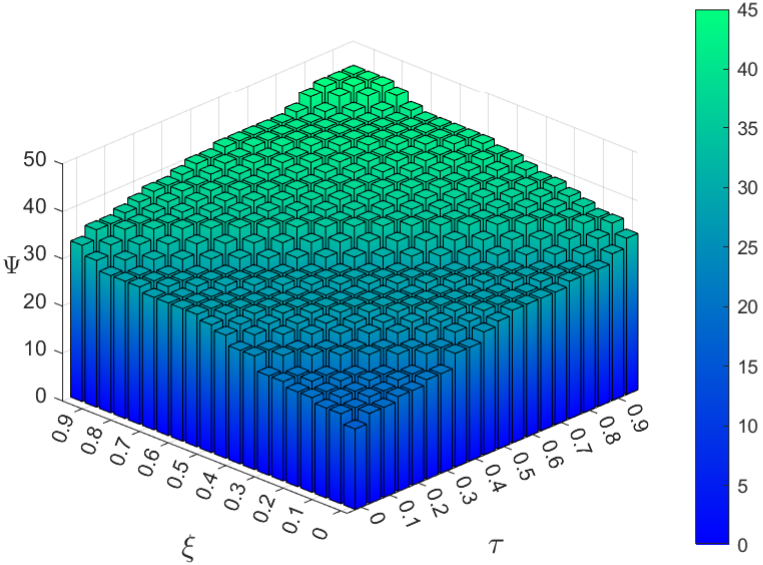}
    \caption{Percentage of vehicles that joined platoon when joint route planner is executed with the Dijkstra's method is employed. The overall average of involvement is 33\%. }
    \label{fig:populationDİjkstra}
\end{figure}

\begin{figure}
    \centering
    \includegraphics[width=\linewidth]{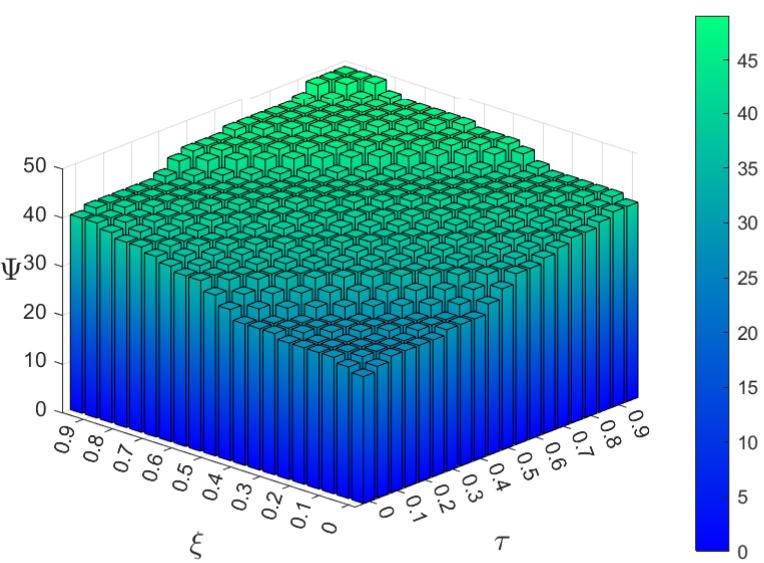}
    \caption{Percentage of vehicles that joined platoon when joint route planner is executed with the A* exploti fatigue as a heuristic. The overall average of involvement is 39\%.}
    \label{fig:populationAstar}
\end{figure}

Fig. \ref{fig:surfPlot} illustrates a comparative analysis of travel cost (in kilometers) between joint and individual route planning methods, plotted as 3D surfaces. The vertical axis represents the travel cost, ranging from approximately 1150 to 1400 KM, while the horizontal axes, labeled $\tau$ and $\xi$, are the percentage of gains within platooning. 
Discrepancies in surface height indicate differences in travel costs.
For $\tau = 100\%$ and $\xi = 18\%$, travel cost improvement of $8\%$ has been achieved. The percentage of vehicles engaged in platooning is visualized in Fig.~\ref{fig:populationDİjkstra} and Fig.~\ref{fig:populationAstar} for Dijkstra's and A* algorithms, respectively. Platooning vehicle involvement rates average $33\%$ when drivers are not fatigued, increasing to $39\%$ when drivers are fatigued.

\section{Conclusion}
\label{Sec:Conc}

% IMPROVE CONCLUSION

%This paper has presented a route optimization system to optimally form a platoon and provided 
%its performance analysis in terms of journey time, fuel consumption and driver fatigue. 
This paper presented a route optimization system based on vehicle collaboration in long journeys by offering a V2X-aided centralized management scheme. It evaluated different cost formulations individually.
The proposed cost formulation considers fuel consumption, travel time, and fatigue level, respectively. While various optimization tools could exploit the proposed solution, the use of Dijkstra and A* algorithms was reported in this paper. Results demonstrate that platooning can improve the energy of the overall string, although a single vehicle increases travel distance. 
Another point of discussion is whether the reduced driver fatigue experienced during platooning could warrant incorporating travel time as a consideration for homologation adjustments.
%The results have demonstrated that joint travel route optimization improves overall travel cost thereby enhancing the feasibility of scheduled platooning, despite individual driving distances being shorter.     
The study will be further improved by validating the solution on a real map at several locations, integrating the analyzed cost functions to derive an accurate cost function, and handling case D. 

\bibliography{ifacconf}             

\begin{thebibliography}{24}
\providecommand{\natexlab}[1]{#1}
\providecommand{\url}[1]{\texttt{#1}}
\providecommand{\urlprefix}{URL }
\expandafter\ifx\csname urlstyle\endcsname\relax
  \providecommand{\doi}[1]{doi:\discretionary{}{}{}#1}\else
  \providecommand{\doi}{doi:\discretionary{}{}{}\begingroup \urlstyle{rm}\Url}\fi

\bibitem[{Adas et~al.(2024)Adas, Barbieri, Morri, Mentasti, Awasthi, Arrigoni, Sabbioni, and Nicoli}]{V2X_akif}
Adas, A., Barbieri, L., Morri, P., Mentasti, S., Awasthi, S., Arrigoni, S., Sabbioni, E., and Nicoli, M. (2024).
\newblock Cooperative {LiDAR}-aided self-localization of {CAVs} in real urban scenario.
\newblock In \emph{16th International Symposium on Advanced Vehicle Control}, 510--516. Springer Nature Switzerland, Cham.

\bibitem[{Bergenhem et~al.(2012)Bergenhem, Shladover, Coelingh, Englund, and Tsugawa}]{platooning_practices}
Bergenhem, C., Shladover, S., Coelingh, E., Englund, C., and Tsugawa, S. (2012).
\newblock Overview of platooning systems.
\newblock In \emph{Proceedings of the 19th ITS World Congress, Oct 22-26, Vienna, Austria (2012)}.

\bibitem[{Bhoopalam et~al.(2018)Bhoopalam, Agatz, and Zuidwijk}]{platooning_eff2}
Bhoopalam, A.K., Agatz, N., and Zuidwijk, R. (2018).
\newblock Planning of truck platoons: A literature review and directions for future research.
\newblock \emph{Transportation research part B: methodological}, 107, 212--228.

\bibitem[{Candra et~al.(2020)Candra, Budiman, and Hartanto}]{dijkstra}
Candra, A., Budiman, M.A., and Hartanto, K. (2020).
\newblock Dijkstra's and {A}-star in finding the shortest path: a tutorial.
\newblock In \emph{2020 International Conference on Data Science, Artificial Intelligence, and Business Analytics (DATABIA)}, 28--32.

\bibitem[{Chavhan et~al.(2023)Chavhan, Kumar, Tiwari, Liang, Lee, and Muhammad}]{plat_csec_3}
Chavhan, S., Kumar, S., Tiwari, P., Liang, X., Lee, I.H., and Muhammad, K. (2023).
\newblock Edge-enabled blockchain-based {V2X} scheme for secure communication within the smart city development.
\newblock \emph{IEEE Internet of Things Journal}, 10(24), 21282--21293.

\bibitem[{Gao et~al.(2023)Gao, Wu, Zhong, and Yau}]{plat_comm_1}
Gao, W., Wu, C., Zhong, L., and Yau, K.L.A. (2023).
\newblock Communication resources management based on spectrum sensing for vehicle platooning.
\newblock \emph{IEEE Transactions on Intelligent Transportation Systems}, 24(2), 2251--2264.

\bibitem[{Goel(2009)}]{consecutiveDriving}
Goel, A. (2009).
\newblock Vehicle scheduling and routing with drivers' working hours.
\newblock \emph{Transportation Science}, 43(1), 17--26.

\bibitem[{Hart et~al.(1968)Hart, Nilsson, and Raphael}]{Astar}
Hart, P.E., Nilsson, N.J., and Raphael, B. (1968).
\newblock A formal basis for the heuristic determination of minimum cost paths.
\newblock \emph{IEEE Transactions on Systems Science and Cybernetics}, 4(2), 100--107.
\newblock \doi{10.1109/TSSC.1968.300136}.

\bibitem[{Heinovski and Dressler(2024)}]{platoon_formation}
Heinovski, J. and Dressler, F. (2024).
\newblock Where to decide? centralized versus distributed vehicle assignment for platoon formation.
\newblock \emph{IEEE Transactions on Intelligent Transportation Systems}, 25(11), 17317--17334.

\bibitem[{Lai et~al.(2021)Lai, Li, and Zheng}]{plat_comm_2}
Lai, C., Li, G., and Zheng, D. (2021).
\newblock {SPSC}: A secure and privacy-preserving autonomous platoon setup and communication scheme.
\newblock \emph{Transactions on Emerging Telecommunications Technologies}, 32(9), e3982.

\bibitem[{Lammert et~al.(2014)Lammert, Duran, Diez, Burton, and Nicholson}]{plat_fuel_consumption}
Lammert, M., Duran, A., Diez, J., Burton, K., and Nicholson, A. (2014).
\newblock Effect of platooning on fuel consumption of class 8 vehicles over a range of speeds, following distances, and mass.
\newblock \emph{SAE International Journal of Commercial Vehicles}, 7, 626--639.

\bibitem[{Liu et~al.(2017)Liu, Ma, Weimerskirch, and Zhu}]{plat_csec_2}
Liu, J., Ma, D., Weimerskirch, A., and Zhu, H. (2017).
\newblock A functional co-design towards safe and secure vehicle platooning.
\newblock In \emph{Proceedings of the 3rd ACM Workshop on Cyber-Physical System Security}, CPSS '17, 81–90. Association for Computing Machinery, New York, NY, USA.

\bibitem[{MacLean et~al.(2003)MacLean, Davies, and Thiele}]{fatigueRelatedAccidents}
MacLean, A.W., Davies, D.R., and Thiele, K. (2003).
\newblock The hazards and prevention of driving while sleepy.
\newblock \emph{Sleep Medicine Reviews}, 7(6), 507--521.

\bibitem[{Malao et~al.(2021)Malao, Pawar, Bhamre, Tagare, and Sorate}]{platooning_def}
Malao, A., Pawar, A., Bhamre, H., Tagare, P., and Sorate, P.G. (2021).
\newblock Vehicle platoon formation and management.

\bibitem[{{On-Road Automated Driving (ORAD) Committee}(2021)}]{SAE}
{On-Road Automated Driving (ORAD) Committee} (2021).
\newblock \emph{Taxonomy and definitions for terms related to driving automation systems for on-zroad motor vehicles}.
\newblock SAE international.

\bibitem[{Santini et~al.(2019)Santini, Salvi, Valente, Pescapè, Segata, and Cigno}]{plat_man_4}
Santini, S., Salvi, A., Valente, A.S., Pescapè, A., Segata, M., and Cigno, R.L. (2019).
\newblock Platooning maneuvers in vehicular networks: A distributed and consensus-based approach.
\newblock \emph{IEEE Transactions on Intelligent Vehicles}, 4(1), 59--72.

\bibitem[{Schmeitz et~al.(2023)Schmeitz, Willemsen, and Ellwanger}]{10132885}
Schmeitz, A.J.C., Willemsen, D.M.C., and Ellwanger, S. (2023).
\newblock {EU ENSEMBLE} project: Reference design and implementation of the platooning support function.
\newblock \emph{IEEE Transactions on Intelligent Transportation Systems}, 24(9), 8988--9003.
\newblock \doi{10.1109/TITS.2023.3276482}.

\bibitem[{Shah and Khattak(2013)}]{ha2}
Shah, S.A.R. and Khattak, A. (2013).
\newblock Road traffic accident analysis of motorways in pakistan.
\newblock \emph{International journal of engineering research and technology}, 2.

\bibitem[{Simić et~al.(2016)Simić, Kocić, Bjelica, and Milošević}]{driver_monitoring}
Simić, A., Kocić, O., Bjelica, M.Z., and Milošević, M. (2016).
\newblock Driver monitoring algorithm for advanced driver assistance systems.
\newblock In \emph{2016 24th Telecommunications Forum (TELFOR)}, 1--4.

\bibitem[{Sivanandham and Gajanand(2020)}]{platooning_eff1}
Sivanandham, S. and Gajanand, M.S. (2020).
\newblock Platooning for sustainable freight transportation: an adoptable practice in the near future?
\newblock \emph{Transport Reviews}, 40(5), 581--606.

\bibitem[{Viterbo et~al.(2024)Viterbo, Brambilla, Cerutti, Specchia, Franceschini, Carambia, Savaresi, and Nicoli}]{V2X_poli}
Viterbo, R., Brambilla, M., Cerutti, M., Specchia, S., Franceschini, D., Carambia, B., Savaresi, S., and Nicoli, M. (2024).
\newblock Evaluating the {V2X} latency for vehicle positioning: A comparison between {5G-V2X} and {ITS-G5}.
\newblock In \emph{2024 IEEE 8th Forum on Research and Technologies for Society and Industry Innovation (RTSI)}, 271--276.

\bibitem[{Viterbo et~al.(2025)Viterbo, Campolo, Cerutti, Awasthi, Arrigoni, Brambilla, and Nicoli}]{OJITS25}
Viterbo, R., Campolo, F., Cerutti, M., Awasthi, S.S., Arrigoni, S., Brambilla, M., and Nicoli, M. (2025).
\newblock A {5G} roadside infrastructure assisting connected and automated vehicles in vulnerable road user protection.
\newblock \emph{IEEE Open Journal of Intelligent Transportation Systems}, 1--1.

\bibitem[{Ying et~al.(2019)Ying, Ma, and Yi}]{plat_man_3}
Ying, Z., Ma, M., and Yi, L. (2019).
\newblock {BAVPM}: Practical autonomous vehicle platoon management supported by blockchain technique.
\newblock In \emph{2019 4th International Conference on Intelligent Transportation Engineering (ICITE)}, 256--260.

\bibitem[{Zhang et~al.(2019)Zhang, Wang, Wu, and Zhang}]{driverFatigue}
Zhang, Q., Wang, Y., Wu, C., and Zhang, H. (2019).
\newblock Research on maximum driving time based on driving fatigue model from field experiment.
\newblock In \emph{2019 5th International Conference on Transportation Information and Safety (ICTIS)}, 1068--1073.

\end{thebibliography}
\end{document}